\documentstyle[prl,aps,multicol,epsfig]{revtex}

\begin{document}
\draft

\title{Optical conductivity of the half-filled Hubbard chain}

\author{E.~Jeckelmann and F.~Gebhard}
\address{Fachbereich Physik, Philipps-Universit\"at Marburg, D-35032 
Marburg, Germany}
\author{F.H.L.~Essler}
\address{Department of Physics, Warwick University,
Coventry, CV4 7AL, UK}

\maketitle

\begin{abstract}
We combine well-controlled analytical and numerical methods to determine
the optical conductivity of the one-dimensional Mott-Hubbard insulator
at zero temperature. A dynamical density-matrix renormalization group
method provides the entire absorption spectrum for all but very small
coupling strengths. In this limit we calculate the conductivity
analytically using exact field-theoretical methods. Above the Lieb-Wu
gap the conductivity exhibits a characteristic square-root
increase. For small to moderate interactions, a sharp maximum occurs
just above the gap. For larger interactions, another weak feature
becomes visible around the middle of the absorption band. 
\end{abstract}


\begin{multicols}{2}
\narrowtext

In quasi one-dimensional materials like, e.g., 
organic semiconductors~\cite{Farges},
the electron-electron interaction accounts 
for the formation of excitons and, to a substantial degree, for
the optical gap itself~\cite{Dionys}. Therefore,
a microscopic theory of the optical absorption in these materials
requires a detailed understanding of one-dimensional Mott 
insulators~\cite{Mott}.

The paradigm for a one-dimensional Mott insulator
is the half-filled Hubbard model~\cite{Hubbard}
\begin{equation}
\hat{H} 
= -t \sum_{l;\sigma} \left( \hat{c}_{l,\sigma}^+\hat{c}_{l+1,\sigma} 
+ {\rm h.c.}\right) + 
U \sum_{l} \hat{n}_{l,\uparrow}\hat{n}_{l,\downarrow} \, .
\label{Hamiltonian}
\end{equation}
It describes $\pi$~electrons with spin $\sigma=\uparrow,\downarrow$ 
which move from one site~$l$ of the carbon backbone to its neighboring
sites. The lattice spacing is set to unity.
The number of electrons~$N$ equals the
number of lattice sites~$L$.
Using periodic boundary conditions, the kinetic energy
is diagonal in momentum space and gives rise to a cosine band,
$\epsilon(k)=-2t\cos(k)$ of width~$W=4t$.
The electrons' mutual Coulomb repulsion
is mimicked by the purely local Hubbard interaction~$U$.

The model~(\ref{Hamiltonian}) is Bethe-Ansatz 
solvable~\cite{LiebWu}.
The optical gap is given by~\cite{Ovchinnikov}
\begin{equation}
\Delta(U) = \frac{16t^2}{U} \int_{1}^{\infty}
\frac{dy \sqrt{y^2-1}}{\sinh(2\pi t y/U)} \; .
\label{gap}
\end{equation}
For small and large~$U/t$ this gives
$\Delta(U\lesssim 2t) = (8t/\pi) \sqrt{U/t} \exp(-2\pi t/U) $
and $\Delta(U\gtrsim 4t) = U - W + 8\ln(2)t^2/U$.
The optical absorption is proportional 
to the real part of the optical conductivity
which is related to the imaginary part of the current-current correlation
function by
$\sigma_1(\omega>0) ={\rm Im}\{\chi_{jj}(\omega>0)\}/\omega$,
and
\begin{mathletters}%
\label{decomp}
\begin{eqnarray}
\chi_{jj}(\omega>0) 
&=& - \frac{1}{L} \langle 0|\hat{\jmath} 
\frac{1}{E_0-\hat{H}+\hbar\omega+i\eta}\hat{\jmath} |0\rangle 
\label{decompB}\\
&=& - \frac{1}{L} \sum_n 
\frac{ |\langle 0 | \hat{\jmath} | n\rangle|^2}{\hbar\omega -(E_n-E_0) +i\eta}
 \; .
\label{decompC}
\end{eqnarray}\end{mathletters}%
Here, $|0\rangle$ is the ground state, $|n\rangle$ are excited states,
and $E_0$, $E_n$ are their respective energies. 
Although $\eta=0^+$ is infinitesimal,
we may introduce a finite value to broaden our resonances at 
$\hbar\omega=E_n-E_0$.
In momentum space, $\hat{\jmath}=-(2et/\hbar)  \sum_{k;\sigma} \sin(k)
\hat{c}_{k,\sigma}^+\hat{c}_{k,\sigma}$, 
is the current operator.
We set $\hbar = 1$ throughout, and for our numerical results we
use $e = t \equiv  1$ in our figures.

Eq.~(\ref{decomp}) shows why it is so difficult to
calculate optical properties
of Mott-Hubbard insulators. The spectrum~$E_n$ is known exactly
but very little is known about the oscillator strengths 
$T^2_{0,n}=|\langle 0| \hat{\jmath}|n\rangle|^2$, although
it can be shown explicitly that they vanish unless 
$E_n-E_0\geq \Delta$~\cite{Fabian}.
Numerical calculations of $\sigma_1(\omega)$ have been
carried out using exact diagonalizations~\cite{ED}
and quantum Monte Carlo simulations~\cite{QMC}.
Unfortunately, these approaches are seriously limited 
in accuracy or accessible system sizes.
Since the system is a Mott insulator and not a Luttinger liquid metal,
standard bosonization techniques 
cannot be applied to the half-filled Hubbard model~\cite{Giamarchi}.
Therefore, the calculation of the optical conductivity of this model is 
an important yet unsolved problem in theoretical solid-state physics.

In this work we employ the dynamical density-matrix 
renormalization group (DDMRG) method~\cite{KuehnerWhite,Rama1}
to determine the optical conductivity 
of the Hubbard insulator over the entire absorption spectrum.
This numerical technique allows us to obtain $\sigma_1(\omega)$
for all interaction strengths as long as the gap is not 
exponentially small, $U\gtrsim 3t$.
For large interaction strengths, $U\gg t$, we confirm
results obtained in the framework of a $1/U$~expansion~\cite{Marburger}.
In the weak-coupling regime ($\Delta\ll t$), we
calculate $\sigma_{1}(\omega)$
analytically using exact field-theoretical methods.
The analytical results agree well with the DDMRG data for
small $U/t$.

We start our analysis with the large-$U$ limit 
($U/t \rightarrow \infty$),
where a rather simple
band picture emerges~\cite{Marburger}. If we ignore corrections of the
order $t/U$ electron transfers are limited to those processes
which conserve the number of double occupancies. 
Due to spin-charge separation, the oscillator strength~$T^2_{0,n}$
can be written as the convolution of the 
charge and spin contributions. The charge contribution 
follows from a simple band picture: we excite one hole 
in the lower Hubbard band,
$\epsilon_{\rm LHB}(k)=\epsilon(k)$, and one double occupancy 
in the upper Hubbard band,
$\epsilon_{\rm UHB}(k)=U-\epsilon(k)$ (antiparallel bands). 
The total momentum of the two charge
excitations is~$q$, and their energy is~$\omega$.
The spin sector enters the current-current correlation function
via the momentum-dependent ground-state form-factor $g_{q}$. 

\begin{figure}
  \begin{center}
    \epsfig{file=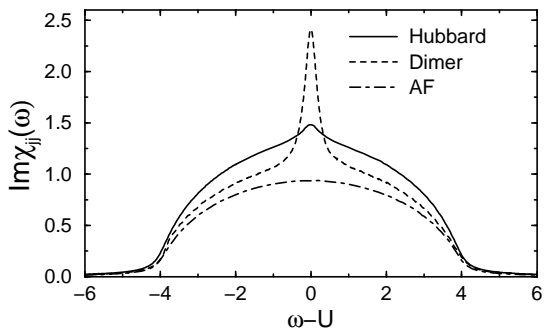,width=7.5cm}
  \end{center}
\caption{Current-current correlation function 
of the large-$U$ Hubbard model 
($U/t \rightarrow \infty$)
and of two related models
with a dimer and a N\'{e}el (AF) spin ground state ($\eta = 0.2$).}
\label{fig:HarrisLange}
\end{figure}

For the large-$U$ Hubbard model itself
($U/t \rightarrow \infty$),
the analysis of $g_q$ is 
rather involved. However, explicit analytical results are available 
for the closely related problems of a dimerized ({\rm DIM}) or 
a N\'{e}el-ordered ({\rm AF})
spin ground state. 
For these states, $g_q$ vanishes unless 
$q=0$ or $q=\pi$,
$g_0^{\rm AF}=2$, $g_{\pi}^{\rm AF}=0$, and 
$g_0^{\rm DIM}=9/4$, $g_{\pi}^{\rm DIM}=1/4$~\cite{Marburger}.
Now, optical transitions occur between two antiparallel bands 
($q=0$; $\epsilon_{\rm LHB}(k)$, $\epsilon_{\rm UHB}(k))$ 
and between two parallel bands ($q=\pi$; 
$\epsilon_{\rm LHB}(k)$, $\epsilon_{\rm UHB}(k+\pi)$).
Then, the optical conductivity becomes
\begin{eqnarray}
\omega \sigma_1(\omega) & =& 
2\pi t^2e^2 \frac{1}{L} \sum_{|k|<\pi} \bigl[
g_\pi \cos^2(k)  \delta(\omega-U)  \nonumber \\
&& + g_0 \sin^2(k) \delta(\omega-U-2\epsilon(k)) \bigr]\label{sigmaHL}
\label{eq:HarrisLange} \\
&=& \pi e^2t^2 \bigl[ g_{\pi} \delta(\omega-U) +
2 g_0/(\pi W^2)   \nonumber \\
&& \left\{\left[\omega-(U-W)\right]\left[(U+W)-\omega\right]\right\}^{1/2}   
\bigr] \nonumber \; .
\end{eqnarray}
For the large-$U$ Hubbard model $g_q$ cannot be calculated analytically. 
We can adopt a ``no-recoil approximation''~\cite{Marburger} to
argue that the dominant contributions to the conductivity again come
from $q=0$ and $q=\pi$. Figure~\ref{fig:HarrisLange} shows DDMRG
results obtained on 64-site lattices for the large-$U$ Hubbard model
and the related models with a dimer or N\'{e}el-ordered spin ground
state.
On the scale of this figure there is no visible difference
between our numerical results for the two models with
spin order and the exact results~(\ref{eq:HarrisLange}) 
with the same broadening $\eta$.
This shows that the DDMRG approach is very accurate in these
limiting cases. 

For the large-$U$ Hubbard model DDMRG data are also in agreement
with eq.~(\ref{eq:HarrisLange})
confirming the no-recoil
approximation in Ref.~\cite{Marburger}. 
In particular, one notes
the small bump in $\sigma_1(\omega)$ at $\omega=U$ 
showing that $g_{\pi} > 0$ (see Fig.~\ref{fig:HarrisLange}). 
This bump is not visible in small chains but becomes more evident
and sharper as we increase the system size.
Therefore, there
are two salient features in the optical absorption of the large-$U$ 
Hubbard model,~(i),~the behavior at threshold is
$\sigma_1(\omega)\sim \sqrt{\omega-\Delta}$, and~(ii), the large density of
states for excitations between parallel bands results in a small but visible 
peak in the middle of the absorption band. 
The second feature is not present in the optical conductivity calculated
using a N\'{e}el-ordered spin ground state~\cite{Marburger,Sumit}
(see also Fig.~\ref{fig:HarrisLange}) because 
the spin form-factor $g_q$ of a N\'{e}el state vanishes for all $q\neq0$. 

The density-matrix renormalization group (DMRG) is known
to be a very accurate numerical method to determine static properties
of low-dimensional lattice systems~\cite{Steve,DMRGbook}.
Recently, K\"{u}hner and White~\cite{KuehnerWhite} developed 
an efficient scheme to calculate dynamical correlation functions 
in Heisenberg spin chains
using DMRG and the correction vector method~\cite{Rama1}.  
We have extended this dynamical DMRG method to correlated 
electron systems, such as the Hubbard model~(\ref{Hamiltonian}).
Our method differs from that of Ref.~\cite{KuehnerWhite} 
in two points: (i)~we calculate correction vectors  
with Ramasesha's algorithm~\cite{Rama2}
instead of a conjugate gradient method, and~(ii)
we compute dynamical correlation functions directly 
from the correction vectors.
    
Our DDMRG method allows us to calculate dynamical
correlation functions, such as the r.h.s.\ of eq.~(\ref{decompB}),
very accurately for fairly large systems ($L\leq 128$)
and a {\sl finite} broadening factor $\eta$.
Thus, our numerical results always correspond to the actual function  
$\sigma_1(\omega)$ convoluted with a Lorentzian
$L(\omega) = \eta/[\pi(\omega^2+\eta^2)]$.
We compared the predictions of our method  
with exact results for a Peierls 
insulator~\cite{Marburger} and the 
two models with a spin-ordered ground state~(\ref{eq:HarrisLange}) 
and found that errors were smaller than 1\% over the entire spectrum.
In the limit of small $U/t$ our numerical results agree very
well with exact field-theory results, see below.
We have also systematically checked various sum rules
relating moments of the function $\sigma_1(\omega)$
to ground-state expectation values that are known exactly or can be
evaluated with great accuracy using a ground-state DMRG
method~\cite{Steve}.
For instance,
\begin{equation}
\int_0^\infty \frac{d\omega}{\pi}
\sigma_1(\omega)=
\frac{t}{2L} \, \langle 0|\sum_{l;\sigma} 
\left( \hat{c}_{l,\sigma}^+\hat{c}_{l+1,\sigma} 
+ {\rm h.c.}\right) |0\rangle\ .
\label{sumr}
\end{equation}
Lastly, the optical gaps deduced from DDMRG data
always agree with the exact result~(\ref{gap})
after taking finite-size corrections into account.

\begin{figure}
  \begin{center}
    \epsfig{file=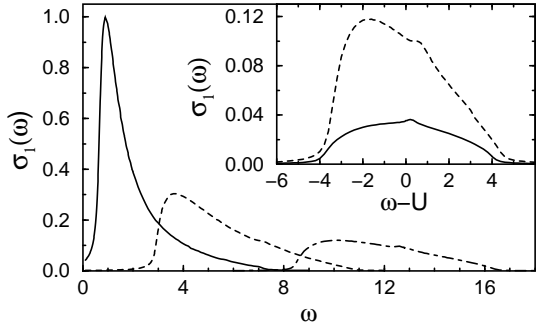,width=7.5cm}
  \end{center}
\caption{Optical conductivity for $U/t = 3,6,12$ (from left to right),
calculated with DDMRG on a 128-site lattice ($\eta=0.1$).
Inset: $\sigma_1(\omega)$ for $U/t = 12$ (dashed) and $40$ 
(solid) calculated on a 64-site chain ($\eta=0.2$).}
\label{fig:Hubbard}
\end{figure}

In Fig.~\ref{fig:Hubbard} we show the optical conductivity
of the Hubbard model calculated with DDMRG for several values of
$U/t$. 
For $U=40t$, $\sigma_1(\omega)$ resembles the
large-$U$ result,
compare Fig.~\ref{fig:HarrisLange}.
As the interaction strength decreases,
the width of the absorption spectrum appears to diminish slightly
from $8t$ for the large-$U$ 
Hubbard model
to less than $7t$ for $U=3t$.
The small peak seen in the large-$U$ Hubbard model 
is barely visible for $U/t < 12$,
but this feature actually subsists at least down to $U=4t$.
As seen in Fig.~\ref{fig:Hubbard}, 
the shape of the optical conductivity changes progressively: 
for strong coupling ($U=12t$), $\sigma_1(\omega)$
displays a broad distribution with a maximum clearly above
the optical absorption threshold $\omega = \Delta$;
for weaker coupling ($U=3t$), 
$\sigma_1(\omega)$ develops a very sharp peak close to the 
threshold and a long tail for higher frequencies. 
For $U/t \rightarrow 0$ this peak eventually
turns into a delta function as all the spectral
weight is concentrated in the Drude peak for $U=0$.

For $U/t \geq 3$ the resolution of our calculations is good enough
to see that the maximum of $\sigma_1(\omega)$ occurs at a frequency
significantly above the optical conductivity threshold.
Our numerical data are compatible with a square-root
increase of the optical conductivity above the Lieb-Wu gap
for all values of $U/t\geq 3$, in agreement with the analytical
result for $U\gg t$. 
Obviously, the exact behavior of $\sigma_1(\omega)$ at the threshold 
cannot be determined from the DDMRG results
alone because the resolution is limited by the finite
broadening~$\eta$ and finite-size effects.
For the same reasons, it is extremely difficult to study the
optical conductivity of the Hubbard chain with DDMRG 
when the optical gap becomes exponentially small ($U/t < 3$). 

For small interaction strengths, $U\ll t$, the half-filled Hubbard
model can be mapped~\cite{affleck89,ezer} onto its low-energy
effective field theory, the SU(2) Thirring model
\begin{eqnarray}
\cal{L}&=&i\bar{\psi}\partial\!\!\!\slash\psi -(U/W)
\sum_{a=1}^3 J_\mu^aJ^{\mu a}\; .
\label{su2thirr} 
\end{eqnarray}
Here, $\psi$ is a doublet of Dirac spinors and 
$J_\mu^a=\frac{1}{2}\bar{\psi}\gamma^\mu\sigma^a\psi$ are SU(2)
currents, where $\gamma^\mu$ are 1+1 dimensional Dirac matrices and
$\sigma^a$ are Pauli matrices. 
The theory~(\ref{su2thirr})
decouples into a massless (spin) and massive (charge) sector.
The charge sector exhibits a SU(2) symmetry 
as is required by the SO(4) symmetry of the half-filled Hubbard
model~\cite{so4}. This symmetry is in general broken in the
Luther-Emery (a.k.a.~$U(1)$ Thirring) model~\cite{LE}, 
and the optical absorption of the half-filled Hubbard model
cannot be calculated using the exact solution at the Luther-Emery
point~\cite{Finkelstein}. 

In the field-theory limit, the current operator is found to be proportional
to a SU(2) current $\hat{\jmath}\propto\int dx\ J^3_1(x)$. This
operator couples only to the gapped charge sector of the theory, so that
only (multi) holon-antiholon scattering states contribute to~(\ref{decompC}). 
It can be shown that lattice and band-curvature effects generate a coupling 
to the spin sector, but this is not important in the field-theory limit. 
We use the spectral representation of the excited states
in terms of scattering states
of holons ($h$), antiholons ($\bar{h}$) and spinons which form a basis of the 
Hilbert space~\cite{smat}.
Using charge-conjugation symmetry one finds that only states with equal
number of holons $N_h$ and antiholons $N_{\bar{h}}$ couple to $J^3_1$.

\begin{figure}
  \begin{center}
    \epsfig{file=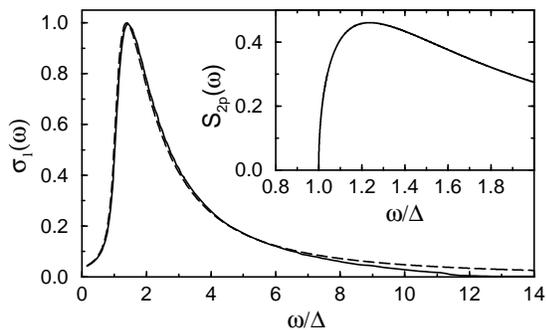,width=7.5cm}
  \end{center}
\caption{Optical conductivity for $U/t=3$ from field theory (dashed)
and DDMRG (solid). 
$\sigma_1(\omega)$
does not vanish below the gap $\Delta = 0.631t$  because
a broadening of $\eta=0.1$ is used.
Inset: universal function $S_{\rm 2p}(\omega)
=\sigma_{\rm 2p}(\omega)/C_2(\Delta)$.}
\label{fig:FTandDDMRG}
\end{figure}

In the field-theory limit the holon/antiholon dispersion 
$E(P)=\sqrt{P^2+(\Delta/2)^2}$ is parameterized in 
terms of a rapidity $\theta$ as
$E(\theta)=(\Delta/2)\cosh\theta$,
$P(\theta)=(\Delta/2)\sinh\theta$.
A scattering state of $N$ (anti)holons with 
rapidities $\{\theta_j\}$ and SU(2) indices $\{\varepsilon_j\}$ 
($\varepsilon_j=h,\bar{h}$) is denoted by
$|\theta_1,\theta_2,\ldots,\theta_N
\rangle_{\varepsilon_1,\varepsilon_2,\ldots \varepsilon_N}$.
Its energy and momentum are $E_N=\sum_{j=1}^NE(\theta_j)$, 
$P_N=\sum_{j=1}^NP(\theta_j)$. 
In this basis, the two-point function of $J^3_1(x)$ 
in the spectral representation reads
\begin{eqnarray}
&&\langle J^3_1(t,x)J^3_1(0,0)\rangle=
\sum_{n=1}^\infty\frac{1}{n!}\sum_{\{\varepsilon_j\}}
\int_{-\infty}^\infty
\prod_{j=1}^n\frac{d\theta_j}{2\pi}\nonumber\\
&&\times\exp\left(-itE_N+ixP_N\right)
|\langle 0|J^3_1(0)|\theta_1,\ldots,\theta_n
\rangle_{\varepsilon_1\ldots\varepsilon_n}|^2.
\label{2pt}
\end{eqnarray}
The matrix elements $\langle 0|J^3_1(0)|\theta_1,\ldots,\theta_n
\rangle_{\varepsilon_1\ldots\varepsilon_n}$
have been determined in \cite{ks}, so that we can evaluate~(\ref{2pt}).
An intermediate state with $N_h$ holons contributes to $\sigma_1(\omega)$ 
only if
$\omega > N_h \Delta$. Taking into account only intermediate states with 
$N_h=N_{\bar{h}}=1$
we obtain ($\nu=\omega/\Delta$)
\begin{eqnarray}
\sigma_{\rm 2p}(\omega) &=& e^2 C_2(\Delta)
\frac{2}{\pi}\frac{\sqrt{\nu^2-1}}{\nu^2}
\Theta(\nu-1)\label{2part}\\
&&
\exp\left(-\int_0^\infty\frac{dx}{x}\frac{1-\cos(x\theta/\pi)\cosh x
}{
\exp(x/2)\cosh(x/2)\sinh x}\right) \ ,
\nonumber
\end{eqnarray}
where $\theta=2 {\rm arccosh} (\nu)$. Formula~(\ref{2part})
is exact in the interval $\Delta\leq \omega\leq 2\Delta$. 
For $\omega >2\Delta$ there are corrections to~(\ref{2part}), which 
are due to multi holon/antiholon states and have a more complicated
structure, but can be shown to be important only at energies $\omega
\gg \Delta$ \cite{CET}.
At present, the normalization $C_2(\Delta)$ 
cannot be calculated analytically. We note that 
$S_{\rm 2p}(\omega)\equiv \sigma_{\rm 2p}(\omega)/C_2(\Delta)$ is a 
universal function of $\nu=\omega/\Delta$.
$S_{\rm 2p}(\Delta\leq \omega \leq 2\Delta)$
is shown in the inset of Fig.~\ref{fig:FTandDDMRG}.

$\sigma_{\rm 2p}(\omega)$ goes through a maximum for 
$\omega \approx 1.24 \Delta$.
For $\omega\to \Delta$ we find
the expected square-root behavior, $\sigma_1(\omega\to \Delta) 
= \sigma_{\rm 2p}(\omega\to \Delta) \sim \sqrt{\omega-\Delta}$.
For the Peierls insulator~\cite{Marburger}, 
a square-root {\sl divergence} occurs, reflecting the
divergence in the density of states for
excitations near the gap. For the Mott-Hubbard insulator,
however, this divergence is suppressed by the 
momentum dependence of the oscillator strengths.
For large $U/t$, eq.~(\ref{sigmaHL}) shows explicitly
that $T^2_{0,n}(|k|\to\pi,\omega\gtrsim \Delta) 
\sim \sin^2(k) $ vanishes quadratically in the vicinity of the gap.

To estimate the normalization $C_2(\Delta)$, we use
the exact sum rule~(\ref{sumr}).
In the weak-coupling limit $U \ll t$, most of the optical weight 
of the Hubbard model~(\ref{Hamiltonian})
must be concentrated at low energy $\omega \sim \Delta$ 
as $\sigma_1(\omega)$ reduces to a single Drude peak
for $U=0$. 
Therefore, in the field-theory limit,
the l.h.s.~of~(\ref{sumr}) can be determined 
from~(\ref{2part}) as the omitted terms have a negligible contribution
at low energy.
For the r.h.s.~we use the exact result at $U=0$.
We obtain $C_2(\Delta) = 1.867 \, t/\Delta$.

In Fig.~\ref{fig:FTandDDMRG} we compare the field-theory
prediction~(\ref{2part}), 
using the exact value $\Delta \approx 0.631 t$ 
of the optical gap (\ref{gap}) rather than its
field theory value, and the DDMRG
result for $U=3t$. We see that both results agree well up to $\omega
\approx 6\Delta$. This is surprising as the field theory is expected
to work only as long as $\Delta\ll t$. 
It appears that even for $U/t=3$, the contribution of states with more
than one holon-antiholon pair to $\sigma_1(\omega)$ is almost
negligible in this range of frequencies. 
If we do the same comparison for larger $U/t$, we see that the range
of $\omega/\Delta$ for which $\sigma_{\rm 2p}(\omega)$ matches the
DDMRG data becomes smaller, but both methods coincide around the
optical absorption threshold even for fairly strong couplings ($U
=6t$). We have no explanation for this universal behavior of the 
optical conductivity above $\Delta$.

In conclusion, we have determined the 
optical conductivity~$\sigma_{1}(\omega)$ of the
one-dimensional Mott-Hubbard insulator over the entire absorption
spectrum for all values of the interaction strength. 
As also reported in~\cite{Carmelo},
we have found a square-root increase of $\sigma_1(\omega)$ at the 
absorption edge in contrast to the square-root divergence
for the one-dimensional Peierls insulator.
In principle, this difference could
be used to distinguish experimentally between the two types of
insulators. However, other interactions, e.g., the long-range parts
of the Coulomb interaction, lattice structure, interchain couplings,
and disorder, can significantly change the absorption spectrum. 
We think that both dynamical-DMRG and field-theory approaches
will permit the reliable calculation of dynamical
properties for more general Hamiltonians which include these additional
interactions.

We gratefully acknowledge helpful discussions with D.~Baeriswyl, 
T.~K\"{u}hner, A.M.~Tsvelik, and S.R.~White.

\end{multicols}

\end{document}